# Forced ignition of laminar premixed spherical flames in evaporating fuel droplet mists

Qiang Li[1], Huangwei Zhang*[1]
[1]Department of Mechanical Engineering, National University of Singapore, 9 Engineering Drive 1, Singapore 117576, Republic of Singapore

**Abstract**
This study aims to build a theoretical model to describe the ignition process of spherical spray flame laden with fuel droplets under overall fuel-lean condition. Two characteristic fronts are introduced to reveal where the evaporation starts and ends in the ignition process. The results demonstrate that the flame has been strengthened with extra fuel droplet adding while the minimum ignition energy (MIE) is decreased. The evolution of the two evaporation fronts in the flame kernel and propagating branches has also be revealed coupled with the evaporation zone length.

**Keywords**
Ignition; Spray flame; Fuel droplet; Evaporation fronts.

**Introduction and physical model**
Spray combustion are widely used in propulsion systems, like aero-engines and rocket engines. Compared to gaseous flame, the existence of fuel/oxidizer sprays in the combustion domain may bring some intriguing features in the ignition process, i.e., the liquid phase interaction with the gaseous flame kernel during the ignition period [1]. Theoretical analysis could provide a good reference on the ignition characteristics of spray flames.
In this work, forced ignition of one-dimensional spherical flame in partially pre-vaporized fuel sprays will be studied. Initially, the gaseous fresh mixture is assumed to be fuel-lean and the fuel droplets are uniformly dispersed. The forced ignition is modelled as a localized energy deposition in fuel sprays. If the ignition energy is larger than a Minimum Ignition Energy (MIE), a flame kernel is generated and continuously propagate outwardly. During the flame development process, based on the droplet distribution relative to the reaction front, two general scenarios exist, i.e., evaporating droplets in: (1) both pre- and post-flame zones and (2) pre-flame zone only. The sketches of their physical models are shown in Figs. 1(a) and 1(b), respectively. There are three characteristic locations for liquid and gas phases, including flame front ($R_f$), droplet evaporation onset ($R_v$) and completion ($R_c$) fronts. For the latter two, $R_v$ corresponds to the location where the droplets are just heated up to boiling temperature and hence start to evaporate. For $R < R_v$, the droplet temperature remains constant and evaporation continues [2–6]. The evaporation onset front $R_v$ is always before the flame front $R_f$, indicating that onset of droplet vaporization spatially precedes the gaseous combustion. Moreover, $R_c$ denotes the location at which all the droplets are critically vaporized. When $R < R_c$, no droplets are left and hence their effects on the gas phase diminish. For brevity, hereafter, we term the first (Fig. 1a) and second (Fig. 1b) cases as heterogeneous (abbreviated as "HT") and homogeneous ("HM") flames, respectively. There are four zones in both flames. Specifically, zone 1 represents the pre-vaporization zone before $R_v$, and 2 indicates pre-flame evaporation zone before $R_f$ for heterogeneous flame and before $R_c$ for homogeneous flame. As for 3, it represents post-flame evaporation zone before $R_c$ for heterogeneous flame, and pre-flame zone without evaporation for homogeneous flame. Meanwhile, 4 is the post-flame droplet-free zone for both flames.

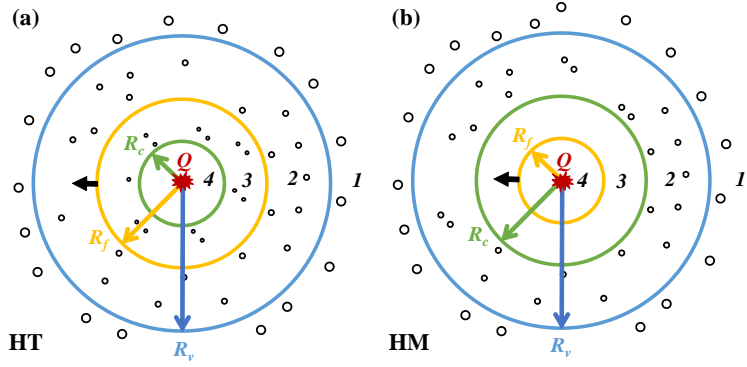

**Figure.1.** Schematic of outwardly propagating spherical flame in liquid fuel mists: (a) heterogeneous flame, (b) homogeneous flame. Circle: fuel droplet. Yellow line: flame front ($R_f$); green line: evaporation completion front ($R_c$); blue line: evaporation onset front ($R_v$). Black arrow: flame propagation direction. Red Spark ($Q$): forced ignition energy at the flame centre.

**Governing equation and boundary condition**
For the gaseous flames, the well-known diffusive-thermal model [7,8] is adopted. This model has been used in numerous studies on gaseous and two-phase flames [2,4–6,9,10]. One-step chemistry is considered, i.e., $F + O \rightarrow P$, with $F$, $O$ and $P$ being fuel vapour, oxidizer and product, respectively. Only fuel-lean fresh mixture (i.e., equivalence ratios of fuel vapor and fuel droplets are above unity) is studied in this work and hence fuel vapour $F$ is the deficient species. Following previous theoretical analysis for both gaseous flames and two-phase flames with dispersed liquid droplets [4,5,9,11,12], we adopt the quasi-steady state assumption in the moving coordinate system attached to the stably propagating flame front $R_f(t)$, i.e. $\eta = r - R_f(t)$. Moreover, due to relatively dilute fuel droplet concentration, their influences on the reaction zone thickness are small and therefore gaseous combustion still dominates [3,5]. In addition, due to the kinematic and thermal equilibrium between the two phases, the droplets approximately follow the carrier gas in velocity and temperature (zone 1)[13,14]. Therefore, the non-dimensional governing equations for gas temperature, fuel vapour mass fraction and fuel droplet loading can be written as

$$-U\frac{dT}{d\eta} = \frac{1}{(\eta + R_f)^2}\frac{d}{d\eta}\left[(\eta + R_f)^2 \frac{dT}{d\eta}\right] + \omega_c - q_v\omega_v \quad (1)$$

$$-U\frac{dY_F}{d\eta} = \frac{Le}{(\eta + R_f)^2}\frac{d}{d\eta}\left[(\eta + R_f)^2 \frac{dY_F}{d\eta}\right] + \omega_c - \omega_v \quad (2)$$

$$-U\frac{dY_d}{d\eta} = -\omega_v \quad (3)$$

The term $\omega_v$ in Eqs. (1)–(3) is the non-dimensional droplet evaporation rate, i.e.,

$$\omega_v = \frac{\Omega(T - T_v)H(T - T_v)}{q_v} \quad (4)$$

Here $H(x)$ is the Heaviside function, which indicates the evaporation process only occurs when the droplet temperature reaches the boiling temperature.
The non-dimensional boundary conditions for both gas phase ($T$ and $Y_F$) and liquid phase ($Y_d$) at the left boundary (spherical centre, $\eta = -R_f$) and the right boundary ($\eta \rightarrow +\infty$) are [2–6]

$$\eta = -R_f: \quad (\eta + R_f)^2 \frac{dT}{d\eta} = -Q, \quad \frac{dY_F}{d\eta} = 0, Y_d = 0 \quad (5)$$

$$\eta \to +\infty: \quad T = 0, \quad Y_F = 1, Y_d = \delta \quad (6)$$

Here $\delta$ is the initial mass loading of the fuel droplet in the fresh mixture. In this paper, the fuel droplet is deemed additive of the gaseous mixture, and therefore the overall fuel mass fraction $1+\delta$ is still under fuel-lean condition. At the evaporation onset front, $\eta = \eta_v$, the gas temperature ($T$), fuel mass fraction ($Y_F$), and fuel droplet mass loading ($Y_d$) satisfy the following jump conditions [2–6]

$$T = T_v, \quad [T] = \left[\frac{dT}{d\eta}\right] = [Y_F] = \left[\frac{dY_F}{d\eta}\right] = 0, Y_d = \delta \quad (7)$$

where the square brackets, i.e., $[f] = f(\eta^+) - f(\eta^-)$, denote the difference between the variables at two sides of a location. At the evaporation completion front, $\eta = \eta_c$, the jump conditions for the gas temperature ($T$), fuel mass fraction ($Y_F$), and droplet mass loading ($Y_d$) take the following form [3]

$$\begin{cases} [T] = \left[\frac{dT}{d\eta}\right] = [Y_F] = \left[\frac{dY_F}{d\eta}\right] = 0, Y_d = 0, & \eta_c > 0 \\ [T] = \left[\frac{dT}{d\eta}\right] = [Y_F] = \left[\frac{dY_F}{d\eta}\right] = 0, [Y_d] = 0, & -R_f < \eta_c < 0 \end{cases} \quad (8)$$

Large activation energy of the gas phase reaction is assumed in this study. The corresponding jump conditions are

$$T = T_f, \quad Y_F = [Y_d] = 0 \quad (9)$$

$$-\left[\frac{dT}{d\eta}\right] = \frac{1}{Le}\left[\frac{dY_F}{d\eta}\right] = [\sigma + (1-\sigma)T_f]^2 \exp\left[\frac{Z}{2}\left(\frac{T_f - 1}{\sigma + (1-\sigma)T_f}\right)\right] = \chi(T_f) \quad (10)$$

**Analytical solution**

Governing equations (1) – (3) with proper boundary and jump conditions (i.e., Eqs. 5 – 9) can be solved analytically. The solutions for gas temperature $T$, fuel vapour mass fraction $Y_F$, and droplet mass loading $Y_d$ in four zones in both homogeneous and heterogeneous flames are derived. For heterogeneous flames, the solutions of $T$, $Y_F$, and $Y_d$ from zones 1 to 4 are (the number subscripts indicate different zones as shown in Fig. 1a)

$$\begin{cases} T_1(\eta) = T_v \frac{I(\eta, U)}{I(\eta_v, U)} \\ T_2(\eta) = T_v + k_1 L_1(\eta) + k_2 L_2(\eta) \\ T_3(\eta) = T_v + \zeta_1 L_1(\eta) + \zeta_2 L_2(\eta) \\ T_4(\eta) = Q[I(\eta_c, U) - I(\eta, U)] + T_v + \zeta_1 L_1(\eta_c) + \zeta_2 L_2(\eta_c) \end{cases} \quad (11)$$

$$\begin{cases} Y_{F,1}(\eta) = 1 + \dfrac{[F_1(\eta_v) - 1]DI(\eta_v, LeU) - DF_1(\eta_v)[I(\eta_v, LeU) - I(0, LeU)]}{DI(\eta_v, LeU)I(0, LeU)} I(\eta, LeU) \\ Y_{F,2}(\eta) = \dfrac{[F_1(\eta_v) - 1]DI(\eta_v, LeU) - DF_1(\eta_v)I(\eta_v, LeU)}{DI(\eta_v, LeU)I(0, LeU)} [I(\eta, LeU) - I(0, LeU)] + F_1(\eta) \\ Y_{F,3}(\eta) = G(\eta) - G(0) \\ Y_{F,4}(\eta) = -G(0) \end{cases} \quad (12)$$

$$\begin{cases} Y_{d,1}(\eta) = \delta \\ Y_{d,2}(\eta) = \delta + \dfrac{\Omega}{Uq_v} \int_{\eta_v}^{\eta} [T_2(\eta) - T_v] d\eta \\ Y_{d,3}(\eta) = \dfrac{\Omega}{Uq_v} \int_{\eta_c}^{\eta} [T_3(\eta) - T_v] d\eta \\ Y_{d,4}(\eta) = 0 \end{cases} \quad (13)$$

where $I(x,y)$, $L_1(\eta)$ and $L_2(\eta)$ are

$$(x,y) = e^{-yR_f} \int_{x}^{+\infty} (\xi + R_f)^{-2} e^{-y\xi} d\xi \quad (14)$$

$$L_1(\eta) = exp\left[-\dfrac{(K+U)(\eta+R_f)}{2}\right] M\left(1 + \dfrac{U}{K}, 2, K(\eta + R_f)\right) \quad (15)$$

$$L_2(\eta) = exp\left[-\dfrac{(K+U)(\eta+R_f)}{2}\right] N\left(1 + \dfrac{U}{K}, 2, K(\eta + R_f)\right) \quad (16)$$

Here $K = \sqrt{U^2 + 4\Omega}$, $M(a,b,c)$ and $N(a,b,c)$ are the Kummer confluent hypergeometric function and the Tricomi confluent hypergeometric function, respectively [15],

$$M(a,b,c) = \dfrac{\Gamma(b)}{\Gamma(b-a)\Gamma(a)} \int_0^1 e^{ct} t^{a-1} (1-t)^{b-a-1} dt \quad (17)$$

$$N(a,b,c) = \dfrac{1}{\Gamma(a)} \int_0^{+\infty} t^{a-1} e^{-ct} (1+t)^{b-a-1} d\xi \quad (18)$$

where $\Gamma(x) = \int_0^{+\infty} t^{x-1} e^{-t} dt$. The expressions for $k_1$, $k_2$, $\zeta_1$, and $\zeta_2$ in Eq. (11) are

$$k_{1,2} = T_v \dfrac{DI(\eta_v, U)}{I(\eta_v, U)} \dfrac{L_{2,1}(\eta_v)}{L'_{1,2}(\eta_v) L_{2,1}(\eta_v) - L'_{2,1}(\eta_v) L_{1,2}(\eta_v)} \quad (19)$$

$$\zeta_{1,2} = \dfrac{(T_f - T_v) L'_{2,1}(\eta_c) + Q L_{2,1}(0) DI(\eta_c, U)}{L'_{2,1}(\eta_c) L_{1,2}(0) - L'_{1,2}(\eta_c) L_{2,1}(0)} \quad (20)$$

where $DI(x,y) = \dfrac{\partial I(x,y)}{\partial x} = -e^{-y(x+R_f)} (x + R_f)^{-2}$. $F_1(\eta)$ and $G(\eta)$ hold as the following form for heterogeneous flames

$$F_1(\eta) = -\dfrac{\Omega}{q_v} \int_0^{\eta} \dfrac{e^{-LeU(x+R_f)}}{(x+R_f)^2} \int_0^x e^{LeU(y+R_f)} [T_2(y) - T_v] (y + R_f)^2 dy dx \quad (21)$$

$$G(\eta) = -\frac{\Omega}{q_v} \int_{\eta_c}^{\eta} \frac{e^{-LeU(x+R_f)}}{(x+R_f)^2} \int_{\eta_c}^{x} e^{LeU(y+R_f)}[T_3(y) - T_v](y+R_f)^2 dy dx \tag{22}$$

$DF_1(\eta)$ is the first order derivative of $F_1(\eta)$.

For homogeneous flames, the solutions of gas temperature $T$, fuel vapor mass fraction $Y_F$, and droplet loading $Y_d$ from zones 1 to 4 are (the number subscripts indicate different zones as shown in Fig. 1b)

$$\begin{cases} T_1(\eta) = T_v \dfrac{I(\eta, U)}{I(\eta_v, U)} \\ T_2(\eta) = T_v + k_1 L_1(\eta) + k_2 L_2(\eta) \\ T_3(\eta) = T_v + \kappa_1 L_1(\eta) + \kappa_2 L_2(\eta) \\ T_4(\eta) = Q[I(\eta_c, U) - I(\eta, U)] + T_v + \kappa_1 L_1(\eta_c) + \kappa_2 L_2(\eta_c) \end{cases} \tag{23}$$

$$\begin{cases} Y_{F,1}(\eta) = 1 + \dfrac{[F_2(\eta_v) - 1]DI(\eta_v, LeU) - DF_2(\eta_v)[I(\eta_v, LeU) - I(0, LeU)]}{DI(\eta_v, LeU)I(0, LeU)} I(\eta, LeU) \\ Y_{F,2}(\eta) = \dfrac{[F_2(\eta_v) - 1]DI(\eta_v, LeU) - DF_2(\eta_v)I(\eta_v, LeU)}{DI(\eta_v, LeU)I(0, LeU)}[I(\eta, LeU) - I(0, LeU)] + F_2(\eta) \\ Y_{F,3}(\eta) = \dfrac{[F_2(\eta_v) - 1]DI(\eta_v, LeU) - DF_2(\eta_v)I(\eta_v, LeU)}{DI(\eta_v, LeU)I(0, LeU)}[I(\eta, LeU) - I(0, LeU)] \\ Y_{F,4}(\eta) = 0 \end{cases} \tag{24}$$

$$\begin{cases} Y_{d,1} = \delta \\ Y_{d,2} = \delta + \dfrac{\Omega}{Uq_v} \int_{\eta_v}^{\eta} (T_2 - T_v) d\eta \\ Y_{d,3} = 0 \\ Y_{d,4} = 0 \end{cases} \tag{25}$$

$F_2(\eta)$ holds as the following form for homogeneous flame,

$$F_2(\eta) = -\frac{\Omega}{q_v} \int_{\eta_c}^{\eta} \frac{e^{-LeU(x+R_f)}}{(x+R_f)^2} \int_{\eta_c}^{x} e^{LeU(y+R_f)}[T_2(y) - T_v](y+R_f)^2 dy dx \tag{26}$$

while $DF_2(\eta)$ is the first order derivative of $F_2(\eta)$.

Moreover, the correlations describing flame speed $U$, flame temperature $T_f$, evaporation onset location $\eta_v$ and droplet completion location $\eta_c$ under different flame radii $R_f$ are also derived.

For HT flame, the correlations are

$$\begin{cases} \zeta_1 L_1'(0) + \zeta_2 L_2'(0) - k_1 L_1'(0) - k_2 L_2'(0) = \chi(T_f) \\ \left\{ \dfrac{[F_1(\eta_v) - 1]DI(\eta_v, LeU) - DF_1(\eta_v)I(\eta_v, LeU)}{DI(\eta_v, LeU)[I(\eta_v, LeU) - I(0, LeU)]} DI(0, LeU) - DG(0) \right\}/Le = \chi(T_f) \\ T_v + k_1 L_1(0) + k_2 L_2(0) = T_f \\ \delta + \dfrac{\Omega}{Uq_v} \int_{\eta_v}^{0} (T_2 - T_v) d\eta = \dfrac{\Omega}{Uq_v} \int_{\eta_c}^{0} (T_3 - T_v) d\eta \end{cases} \tag{27}$$

where $DG(\eta)$ is the first order derivative of $G(\eta)$.

For HM flames, the correlations are

$$\begin{cases} -\left[Q + \dfrac{k_1 L_1'(\eta_c) + k_2 L_2'(\eta_c)}{DI(\eta_c, U)}\right] R_f^{-2} e^{-UR_f} = \chi(T_f) \\ \dfrac{[F_2(\eta_v) - 1]DI(\eta_v, LeU) - DF_2(\eta_v)I(\eta_v, LeU)}{DI(\eta_v, LeU)I(0, LeU)} \dfrac{DI(0, LeU)}{Le} = \chi(T_f) \\ T_v + k_1 L_1(\eta_c) + k_2 L_2(\eta_c) + \dfrac{k_1 L_1'(\eta_c) + k_2 L_2'(\eta_c)}{DI(\eta_c, U)}[I(0, U) - I(\eta_c, U)] = T_f \\ \delta + \dfrac{\Omega}{Uq_v} \int_{\eta_v}^{\eta_c} (T_2 - T_v) d\eta = 0 \end{cases} \quad (28)$$

**Results and discussion**

Analysis on the ignition of spherical spray flame will be conducted based on the correlations derived above (Eqs. 27 and 28). $Z$ and $\sigma$ are set to be $Z = 10$ and $\sigma = 0.15$ [9,13]. $T_v$ and $q_v$ is assumed to be 0.15 and 0.4, respectively [16].

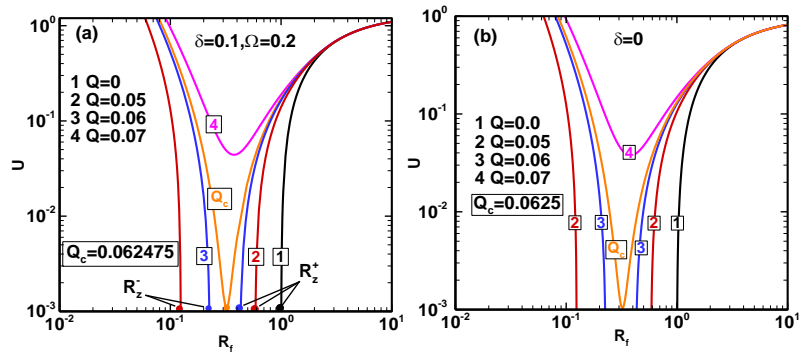

**Figure.2.** Flame propagation speed as a function of flame radius at different ignition energies: (a) $\delta = 0.1$, $\Omega = 0.2$. $Le = 1.0$ and (b) $\delta = 0$ (gaseous flame).

Figure 2(a) shows the flame propagation speed as a function of flame radius at different ignition energies for spray flame. The Lewis number is 1.0. For comparison, the results from gaseous flame is also shown in Fig. 2(b). The trajectories of spray flames are considerably affected by the ignition energy. Specifically, when $Q = 0$, the spray flame is initiated at a flame ball (termed as the outer flame ball, marked as $R_Z^+$) with the radius of about 1.0, and propagate outwardly towards larger radii. With increased ignition energy $Q$, a new flame kernel branch emerges, besides the right flame propagating branch. The flame kernel branches start at a high propagation speed, resulting from the deposited ignition energy. However, due to insufficient energy, the flame kernel propagating speed gradually decays until a flame ball is formed ($U = 0$). This is the inner flame ball $R_Z^-$, corresponding to ignition failure as the flame kernel cannot propagate outwardly. With further increased ignition energy, the flame kernel branch and the flame propagating branch move towards each other, with approaching flame ball radii from them. When the ignition energy is equal to the MIE $Q_c$ (e.g., 0.062475 in Fig. 2a), the twin flame ball solutions coincide, i.e., $R_Z^+ = R_Z^-$, and therefore the two branches merge. For $Q > Q_c$, the flame kernel can continuously propagate outwardly and consecutively reach the spherical flames with large radii.

Comparing Figs. 2(a) and 2(b), we can see that the flame trajectories in spray flames and gaseous flames are similar, indicating that the existence of fuel droplet on overall lean mixture has limited influence on flame ignition process. The small decrease in MIE for lean spray flame could be explained as the chemical heat release of the fuel vapor from droplet evaporation is larger than the evaporative heat loss. However, this decrease is small since the droplet loading is small, while the net heat release from evaporation product is also minimal in the flame kernel branch compared to the ignition energy input. This also leads to a higher propagation speed for spray flame when it propagates at a large flame radius (e.g., line 4 at $R_f$ = 10 in Figs. 2a and 2b).

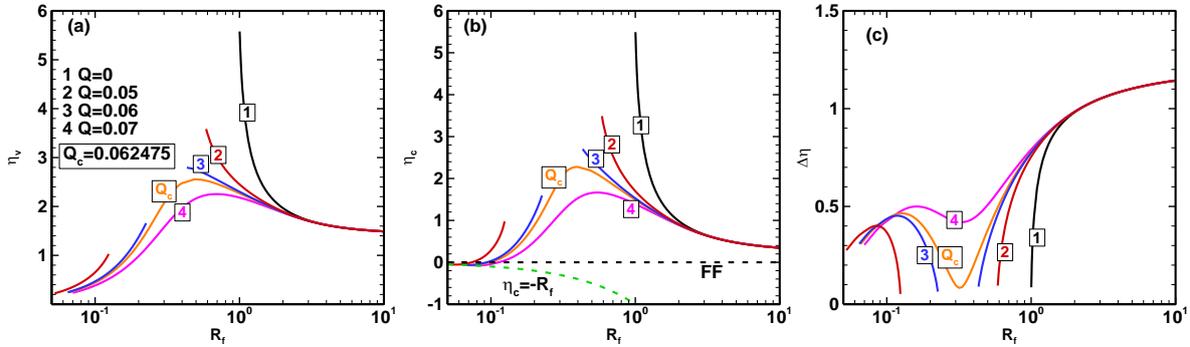

**Figure.3.** (a) Evaporation onset location, (b) evaporation completion location and (c) evaporation zone length as functions of flame radius at different ignition energies when $\delta$ = 0.1, $\Omega$ = 0.2 for $Le$ = 1.0. FF: Flame front.

The novelty of our theoretical model is that it can reveal the evaporation related fronts (onset and completion) within the flame initiation and propagation process. Here we define the length of the droplet evaporation zone, $\Delta \eta = \eta_v - \eta_c$, based on the difference between the droplet onset and completion locations. The changes of the evaporation onset and completion locations and the evaporation zone length are presented in Fig. 3. When the ignition energy $Q$ is less than $Q_c$ (e.g., lines 1–3), the solutions for droplet onset and completion locations have left and right branches. They respectively correspond to the flame kernel and propagating branches. Both droplet evaporation onset front monotonically increases with flame radius at the flame kernel branch. Because the effects of the ignition energy fade, and accordingly the temperature gradient in the pre-heat zone is decreased. For the evaporation completion front, it is clear that at the very beginning moment, the droplet distributed in the whole domain as $\eta_c = -R_f$. Later, with the ignition energy adding, both the flame front and evaporation completion front would move outwardly. However, as the flame propagation speed rapidly decreased due to the fading of ignition energy, the flame front left behind the evaporation completion front. Hence, in the flame kernel branch, the flame would experience a transition from HT to HM. For the flame propagation branch, both evaporation onset and completion locations decrease as the spray flame expands. This is due to the gradual enhancement of the flame.

When the ignition energy equals the MIE (e.g., line 4), the two separate branches for the evaporation onset and completion locations merge, where the inner and outer flame ball solutions coincide. In this scenario, both $\eta_c$ and $\eta_v$ increase due to the flame kernel decaying, and then decrease towards the spherical spray flames with large radii. If the ignition energy is further increased, this non-monotonicity gradually becomes weak since the initial effect of the ignition energy is strengthened.

Figure 3(c) shows the length of the evaporation zone as a function of flame radius. For the flame kernel branch, the evaporation zone length $\Delta \eta$ increases and quickly decreases with the flame kernel decaying process. However, in the flame propagation branch, the evaporation zone length is increased with the flame radius. When the ignition energy is larger than MIE, the evaporation length curves of the two branches merge.

**Conclusions**

A theoretical model has been established to describe the flame initiation process coupled with correlations to depict the nonlinear relation between the flame propagation speed, flame temperature, evaporation onset and completion front. The results on the minimum ignition of lean spray flame laden with fuel droplets shows that the extra fuel droplet addition decreases the MIE. Meanwhile, the results of the evaporation fronts reveal that the flame is initiated from heterogeneous flame to homogeneous flame in the flame kernel branch, regardless of the amount of the ignition energy.

**Acknowledgments**

QL is supported by NUS Research Scholarship. The open source GNU Scientific Library (GSL) is used to do the numerical calculation.

**Nomenclature**

| | | | |
|---|---|---|---|
| $U$ | normalized flame speed | $T$ | normalized gaseous temperature |
| $Y_F$ | normalized fuel mass fraction | $Y_d$ | normalized evaporation rate |
| $\omega_c$ | normalized chemical reaction rate | $\omega_v$ | normalized evaporation rate |
| $R_f$ | normalized flame radius | $Le$ | Lewis number of fuel |
| $\Omega$ | heat transfer coefficient | $\delta$ | initial droplet loading |
| $T_f$ | normalized flame temperature | $T_v$ | normalized boiling temperature |
| $q_v$ | normalized latent heat of vaporization | $\eta$ | flame front attached coordinate |
| $\eta_v$ | evaporation onset front | $\eta_c$ | evaporation completion front |
| $Z$ | Zel'dovich number | $\sigma$ | thermal expansion ratio |